# Two Types of Microwave-Induced Magnetoresistance Oscillations in a 2D Electron Gas at Large Filling Factors

A. A. Bykov*, A. V. Goran, D. R. Islamov, and A. K. Bakarov

*Institute of Semiconductor Physics, 630090 Novosibirsk, Russia*
*e-mail: bykov@thermo.isp.nsc.ru*

**Jing-qiao Zhang, and Sergey Vitkalov**
*Physics Department, City College of the City University of New York, New York 10031, USA*

The influence of microwave radiation (1.2-140 GHz) on resistance of high-mobility two-dimensional electron gas in GaAs quantum wells is studied. Two series of microwave-induced magnetoresistance oscillations periodic in $1/B$ were observed under microwave radiation. The periods of oscillations are determined by the microwave frequency and power, correspondingly. The experimental data is qualitatively explained by photon-assisted transport and Zener tunneling between Landau orbits.

Microwave-induced magnetoresistance oscillations of 2D electron gas in GaAs/AlGaAs heterostructures were recently discovered at large filling factors [1]. It was found that these oscillations are periodic in $1/B$ and oscillation peaks position is determined by $\omega/\omega_c$, where $\omega$ – microwave frequency and $\omega_c$ – cyclotron frequency in magnetic field $B$. The microwave-induced magnetoresistance oscillations in high-mobility 2D electron gas systems were explained in terms of Landau-level transitions between spatially shifted oscillators. This kind of oscillations was theoretically predicted over 30 years ago [2] and its experimental observation is mainly due to advances in technology of fabrication of high-quality modulation doped semiconductor structures.

It was quite recently shown that magnetoresistance oscillations with the period determined by microwave power can be observed in high-mobility electron systems [3], as opposed to [1] where the period is determined by $\omega/\omega_c$. This kind of microwave-induced oscillations at large filling factors can be qualitatively explained by Zener tunneling between tilted Landau levels [4, 5].

In this paper we studied an influence of microwave irradiation on high-mobility 2D electron gas resistance $R_{xx}$ in GaAs quantum wells with AlAs/GaAs superlattice barriers at T = 4.2 K and microwave frequency of (1.2 – 140) GHz in magnetic fields $B$ < 2 T. In such modulation-doped heterostructures we have found microwave induced $R_{xx}$ oscillations periodic in inverse magnetic field $1/B$. We have showed that in the microwave frequency range (1.2–37) GHz the longitudinal resistance $R_{xx}$ of the studied 2D electron systems exhibits oscillations with the position in magnetic field $B$ determined by the microwave power [3]. In the microwave frequency range (37–140) GHz we have found magnetoresistance oscillations whose position in magnetic field $B$ is determined by the microwave frequency, similarly to the oscillations observed in high-mobility GaAs/AlGaAs heterostructures [1]. We found that these two types of the oscillations coexist in our samples at microwave frequency of 37 GHz and temperature 4.2 K.

Our samples were cleaved from the wafers of the high-mobility GaAs quantum wells grown by solid source molecular beam epitaxy on semi-insulating (001) GaAs substrates. The width of the GaAs quantum wells was 13 nm. AlAs/GaAs type-II superlattices served as barriers, which made it possible to obtain a high-mobility 2D electron gas with high electron density. In dark, the electron density and mobility of the 2D electron gas in our samples were $n_e = 1.18 \cdot 10^{16}$ m$^{-2}$ and $\mu = 91$ m$^2$/V s, respectively. After brief light illumination, the electron density and mobility of the 2D electron gas in our samples were $n_e = 1.28 \cdot 10^{16}$ m$^{-2}$ and $\mu = 111$ m$^2$/V s, respectively.

Measurements were carried out at $T$=4.2 K in magnetic field $B$ up to 2 T on 50 μm wide Hall bars with distance of 250 μm between potential contacts. Microwave radiation in frequency region of (1.2-12) GHz was supplied to the sample through a coaxial cable and was fed to the 2D electron gas through current contacts of the Hall bars. Microwave radiation in frequency region of (37-140) GHz was supplied to the sample through a waveguide. The longitudinal resistance was measured using 1 μA current at frequency of (0.3-1) kHz. Five samples are measured. All samples demonstrate the same behavior.

In Fig. 1 we present $R_{xx}(B)$ in the presence of microwave radiation of different power and frequency 140 GHz. It is clearly seen that the amplitude of Shubnikov-de Haas (SdH) oscillations is damped by microwave radiation. From the other hand, microwave radiation produces gigantic resistance oscillations. The analysis of the positions of these oscillations maxima shows that they are periodic in inverse magnetic field. Also, it is seen that there are several areas where the curves almost intersect. Two such areas are determined by the conditions: $\omega = \omega_c$ and $\omega = \omega_c/2$.

The stars in Fig. 1 denote the position of the highest maximum. Its position weakly depends on the microwave power. The weak dependence of the positions of the minima and the maxima in $R_{xx}(B)$ on the microwave power at frequency 140 GHz, as well as the intersection of the curves, produced for different microwave powers, in magnetic field determined by condition $\omega = \omega_c$, indicates that the oscillations shown in Fig. 1 are of the same nature as in [1], and can be explained by indirect inter-Landau-levels transitions, witch involve absorption of microwave quanta and are accompanied by scattering processes that alter the electron momentum [2, 6-14].

This conclusion is in full agreement with the fact that the positions of minima and maxima in magnetic field depends on microwave frequency (Fig. 2). The lowest microwave frequency that allowed us to observe this kind of the oscillations in our samples at temperature 4.2 K was 37 GHz.

In Fig. 3 we present $R_{xx}(B)$ for different power of the microwave radiation of frequency 1.2 GHz. It is seen that microwave-induced resistance oscillations appear at large filling factors, and their positions in magnetic field depend on the microwave power. The position of the highest maximum depends on the microwave power and is denoted by stars. These oscillations are periodic in 1/B similarly to the oscillations observed in microwave radiation of frequency 140 GHz. We have shown [5] that for AC frequency below 100 kHz the position of the AC induced oscillations in magnetic field is proportional to the AC amplitude.

In Fig. 4a we present $R_{xx}(B)$ dependence for two values of microwave power at frequency 37 GHz. Fig. 4b shows a coexistence of two kinds of oscillations at frequency 37 GHz. The position of the minimum and the maximum in magnetic field near $\omega = \omega_c$ (denoted by arrow) depends on the microwave frequency (Fig. 2) and does not depends on the power (Fig. 1), while the position of the maxima denoted by stars (Fig. 3 and Fig. 4) move to higher magnetic fields with the power increasing.

In conclusion, we have shown the coexistence of two kinds of radiation induced $R_{xx}(B)$ oscillations in high-mobility 2D electron systems.

This work was supported by the Russian Foundation for Basic Research (project no. 04-02-16789), INTAS (grant no. 03-51-6453), and NSF Grant Nos. DMR 0349049, DOEFG02-84-ER45153.

[1] M. A. Zudov, R. R. Du, J. A. Simmons, and J. L. Reno, Phys. Rev. B **64**, 201311(R) (2001).
[2] V. I. Ryzhii, Fiz. Tverd. Tela (Leningrad) **11**, 2577 (1969) [Sov. Phys. Solid State **11**, 2078 (1970)].
[3] A. A. Bykov, A. K. Bakarov, A. K. Kalagin, and A. I. Toropov, Pis'ma Zh. Eksp. Teor. Fiz. **81**, 348 (2005) [JETP Lett. **81**, 284 (2005)].
[4] C. L. Yang, J. Zhang, R. R. Du, J. A. Simmons, and J. L. Reno, Phys. Rev. Lett. **89**, 076801 (2002).
[5] A. A. Bykov, Jing-qiao Zhang, Sergey Vitkalov, A. K. Kalagin, and A. K. Bakarov, Phys. Rev. B **72**, 245307 (2005).
[6] V. I. Ryzhii, R. A. Suris, and B. S. Shchamkhalova, Fiz. Tekh. Poluprovodn. (S.-Peterburg) **20**, 2078 (1986) [Sov. Phys. Semicond. **20**, 1299 (1986)].
[7] A. C. Durst, S. Sachdev, N. Read, and S. M. Girvin, Phys. Rev. Lett. **91**, 086803 (2003).
[8] V. Ryzhii, R. Suris, J. Phys.: Condens. Matter **15**, 6855 (2003).
[9] V. Ryzhii and V. Vyurkov, Phys. Rev. B **68**, 165406 (2003).
[10] V. Ryzhii, Phys. Rev. B **68**, 193402 (2003).
[11] X. L. Lei and S. Y. Liu, Phys. Rev. Lett. **91**, 226805 (2003).
[12] V. B. Shikin, Pis'ma Zh. Eksp. Teor. Fiz. **77**, 281 (2003) [JETP Lett. **77**, 236 (2003)].
[13] M. G. Vavilov and I. L. Aleiner, Phys. Rev. B **69**, 035303 (2004).
[14] V. Ryzhii, A. Chaplik, and R. Suris, Pis'ma Zh. Eksp. Teor. Fiz. **80**, 412 (2004) [JETP Lett. **80**, 363 (2004)].

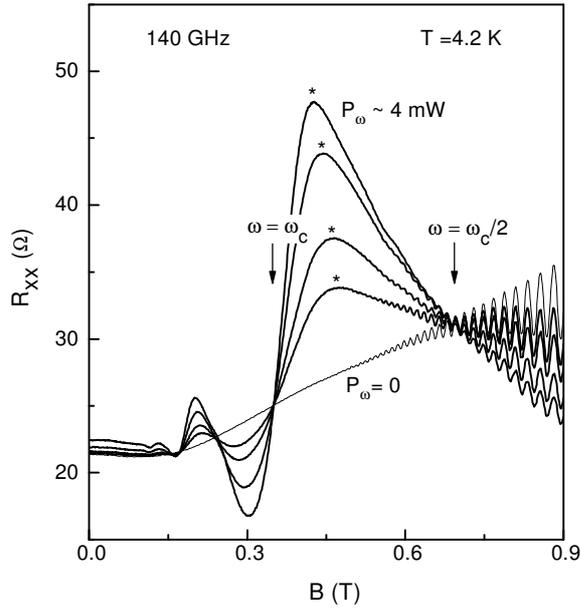

**Fig. 1.** Magnetoresistance $R_{xx}$ with microwave (140 GHz) illumination on (thick line) and off (thin line) at $T = 4.2$ K for different levels of the oscillator output powers $P_\omega$. The asterisks mark the highest maximum.

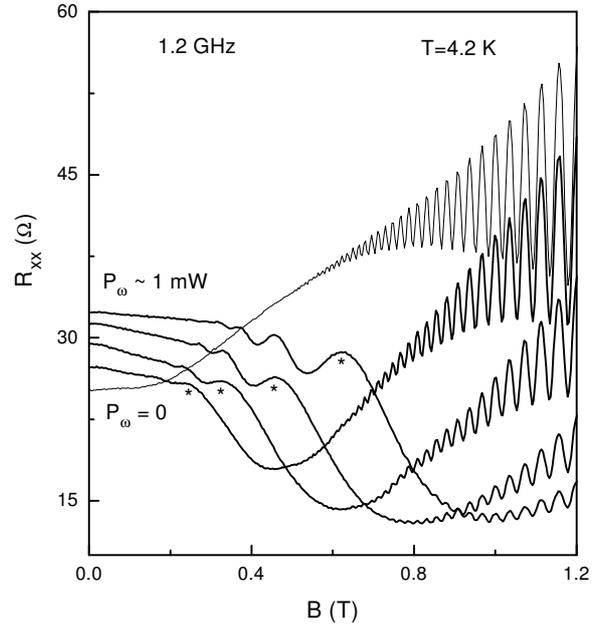

**Fig. 3.** Magnetoresistance $R_{xx}$ with microwave (1.2 GHz) illumination on (thick line) and off (thin line) at $T = 4.2$ K for different levels of the oscillator output powers $P_\omega$. The asterisks mark the highest maximum.

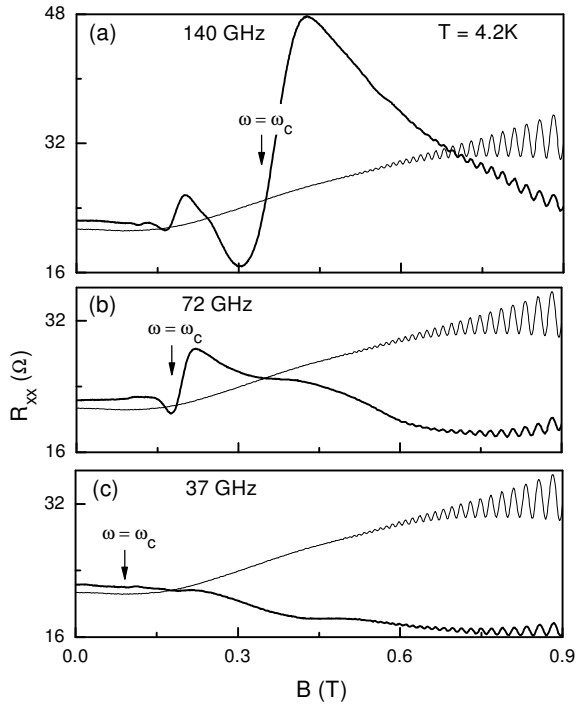

**Fig. 2.** Magnetoresistance $R_{xx}$ with microwave illumination on (thick line) and off (thin line) at $T = 4.2$ K for selected frequencies. The positions of the cyclotron resonance are marked by arrows.

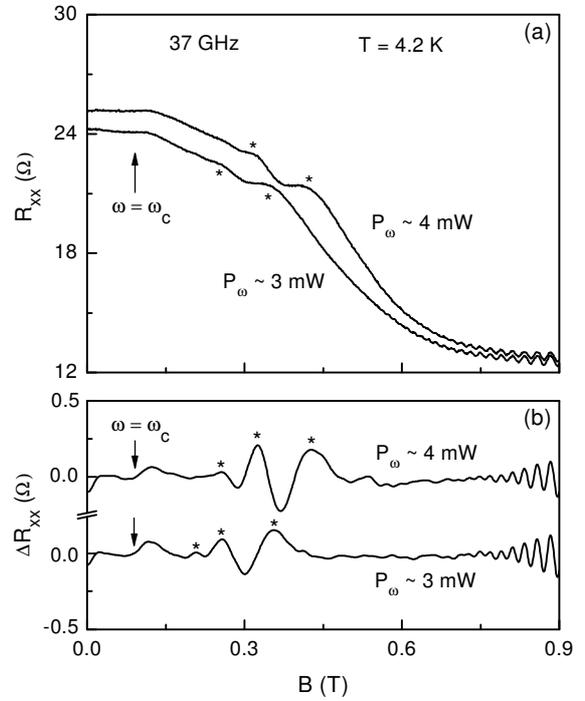

**Fig. 4.** (a) Magnetoresistance $R_{xx}$ with microwave (37 GHz) illumination on at $T = 4.2$ K for two different levels of the oscillator output powers $P_\omega$. The asterisks mark the maxima. (b) Dependences with subtracted monotonic component.